\documentclass[preprint,twoside,11pt]{article}

\usepackage{blindtext}

\usepackage[abbrvbib, preprint]{dmlr2e}

\usepackage{dmlr2e}
\usepackage{cleveref}  
\usepackage[table]{xcolor}
\usepackage{tabularx}
\usepackage{colortbl}

\usepackage{lastpage}
\dmlrheading{25}{2025}{1-\pageref{LastPage}}{5/25}{}{21-0000}{Aaron Green, Zihan Nie, Hanzhen Qin, Oshani Seneviratne, and Kristin P. Bennett} 

\ShortHeadings{FinSurvival: Survival Modeling for Finance}{Green, Nie, Qin, Seneviratne, Bennett}
\firstpageno{1}

\begin{document}

\title{FinSurvival: A Suite of Large Scale Survival Modeling Tasks from Finance}

\author{\name Aaron Green \email greena12@rpi.edu \\
       \addr Department of Mathematical Sciences\\
       Rensselaer Polytechnic Institute\\
       Troy, NY 12180-3950, USA
       \AND
       \name Zihan Nie \email niez@rpi.edu \\
       \addr Department of Mathematical Sciences \\
       Rensselaer Polytechnic Institute \\
       Troy, NY 12180-3950, USA
       \AND
       \name Hanzhen Qin \email qinh2@rpi.edu \\
       \addr Department of Computer Science \\
       Rensselaer Polytechnic Institute \\
       Troy, NY 12180-3950, USA
       \AND Oshani Seneviratne \email senevo@rpi.edu \\
       \addr Department of Computer Science \\
       Rensselaer Polytechnic Institute \\
       Troy, NY 12180-3950, USA
       \AND Kristin P. Bennett \email bennek@rpi.edu \\
       \addr Departments of Mathematical Sciences  and Computer Science\\
       Rensselaer Polytechnic Institute \\
       Troy, NY 12180-3950, USA
       }


\maketitle

\begin{abstract}
    Survival modeling predicts the time until an event occurs and is widely used in risk analysis; for example, it's used in medicine to predict the survival of a patient based on censored data.  
    There is a need for large-scale, realistic, and freely available datasets for benchmarking artificial intelligence (AI) survival models. In this paper, we derive a suite of 16 survival modeling tasks from publicly available transaction data generated by lending of cryptocurrencies in Decentralized Finance (DeFi). Each task was constructed using an automated pipeline based on choices of index and outcome events. For example, the model predicts the time from when a user borrows cryptocurrency coins (index event) until their first repayment (outcome event). We formulate a survival benchmark consisting of a suite of 16 survival-time prediction tasks (FinSurvival). We also automatically create 16 corresponding classification problems for each task by thresholding the survival time using the restricted mean survival time. With over 7.5 million records,  FinSurvival provides a suite of realistic financial modeling tasks that will spur future AI survival modeling research. Our evaluation indicated that these are challenging tasks that are not well addressed by existing methods. FinSurvival enables the evaluation of AI survival models applicable to traditional finance, industry, medicine, and commerce, which is currently hindered by the lack of large public datasets. Our benchmark demonstrates how AI models could assess opportunities and risks in DeFi.  In the future, the FinSurvival benchmark pipeline can be used to create new benchmarks by incorporating more DeFi transactions and protocols as the use of cryptocurrency grows.
\end{abstract}


\section{Introduction}

        Survival data, also called time-to-event data, is used to create models for how long it takes for certain events to occur. This kind of data arises in a wide range of disciplines, most notably in finance, where events of interest could include loan defaults~[\cite{duffieFrailty, landoCreditRisk}], bankruptcy~[\cite{sumwayHazardModels}], or customer churn, and in medicine, where events of interest could include the recovery or death of a patient.  Given the nature of these disciplines, survival datasets about these events can be difficult to obtain. These datasets tend to be sensitive or private, with many deep-learning-based survival methods being based on economic data that requires expensive paid subscriptions or medical datasets that have restricted or no availability~[\cite{ranganath2016deep, miscouridou2018deep, jing2019deep, deephitPaper}]. In addition, these datasets tend to be limited in size. Many popular survival datasets such as METABRIC~[\cite{metabricPaper}] or SUPPORT~[\cite{Knaus1995TheSP}] have only between 1,500 and 10,000 records, and in SurvSet~[\cite{drysdale2022survset}], which contains a repository of 76 survival datasets, the largest dataset has just 52,422 records, and most are much smaller. Not only do these datasets have relatively few records, but they also have very few features, most containing fewer than ten features. One prominent application of survival analysis is in the field of Omics. The data in this domain typically contains huge amounts of features ($>$ 4000), but very few records ($<$ 100). Paid economic survival data can contain large datasets such as Moody's Default and Recovery Database~\footnote{\url{https://www.moodys.com/sites/products/ProductAttachments/DRD Documentation v2/DRDV2_FAQ.pdf}}, which has over 850,000 records, but the cost to access the data can be prohibitively high.         
        
        Given the effectiveness of deep learning models in nearly every discipline, and given these models' need for lots of training data, the existing survival datasets are too small to truly assess the capacities of state-of-the-art models. We address this gap by publishing a novel collection of survival datasets based on free, publicly accessible financial transaction data from the decentralized finance (DeFi) space that consists of 16 different time-to-event scenarios and combines for 7,698,497 records, averaging over 481,000 records per dataset. We show a comparison of our data size with various public survival datasets from other domains in \cref{tab:datasetComparison}. To the best of our knowledge, this is the first large-scale, publicly available financial survival dataset derived from DeFi transactions. Additionally, our datasets contain no personal identification information (PII) nor intellectual property (IP) and are freely available for research use.

        \begin{table}
            \caption{Comparison of our dataset with other publicly available survival datasets.}
            \rowcolors{1}{}{lightgray!25}
            \begin{tabularx}{\linewidth}{XXXXX}
                \hline
                Dataset & Domain & \# Records & \# Features & Source \\
                \hline
                FinSurvival & Finance & 7,698,497 & 128 & this paper \\
                Melanoma & Omics & 41 & 642 & \cite{melanoma}\\
                Ovarian & Omics & 58 & 19,818 & \cite{ovarianData} \\
                SUPPORT & Clinical & 9,105 & 47 & \cite{Knaus1995TheSP}\\
                METABRIC & Clinical &  1,980 & 9 & \cite{metabricPaper}\\
                WHAS & Clinical & 1,638 & 5 & \cite{Floyd2009-dg} \\
                GBSG & Clinical & 686 & 9 & \cite{gbsg} \\
                hdfail & Engineering & 54,422 & 6 & \cite{hdfailPackage} \\
                \hline
            \end{tabularx}
            \label{tab:datasetComparison}
        \end{table}

        DeFi is an emerging area within the cryptocurrency world that aims to provide financial services without the need for traditional banks. It uses blockchain technology and smart contracts to offer services like lending, borrowing, trading, and earning interest on crypto assets. One major type of DeFi application is the lending protocol. Lending protocols function similarly to banks in traditional finance, allowing users to deposit their monetary assets into a savings account and accrue some interest, as well as borrow funds from the protocol using their deposited assets as collateral. In this work, we use data from one of the leading lending protocols, Aave~[\cite{aave-whitepaper}], which has more than \$27 billion locked across eight different networks and 15 markets as of April 4, 2025.

        In Aave, we study five key transaction types that facilitate lending and borrowing activities:  deposits, borrows, repays, withdraws, and liquidations. The protocol {\em users} are actually cryptocurrency wallets with no identifying information.   {\em Deposit} transactions involve users supplying a cryptocurrency to the protocol to earn interest over time while providing liquidity for borrowers.  Deposits also serve as collateral for the users' loans.    {\em Borrow} transactions allow users to take out loans against their deposited collateral, enabling them to access liquidity without selling their assets. {\em Repay} transactions refer to the act of paying back borrowed funds, reducing the borrower's outstanding debt and interest obligations. {\em Withdraw} transactions enable users to retrieve their deposited assets, provided they still meet the collateral requirements after the withdrawal. Lastly, {\em liquidation} transactions occur when a borrower's collateral value falls below the required threshold, triggering the sale of collateral to repay the loan and protect the protocol's solvency. The cryptocurrency used in a transaction is referred to as the {\em reserve}. 
        These transactions collectively define the fundamental financial dynamics of lending in Aave.  
        There are some natural questions one can ask about user behaviors based on these transaction types, such as ``How long do users take to repay loans after borrowing?'' or ``How long do users leave deposited money in their account before withdrawing it?'' We build distinct survival datasets to model these types of questions.

        In this paper, we collected raw Aave user transaction data from TheGraph\footnote{\url{thegraph.com}} and created a pipeline for transforming the transaction data into survival data. This process involves selecting one of the transaction types as an index event and another transaction type as an outcome event, collating transactions based on the user and coin (i.e., reserve) used in the transaction, and computing the time elapsed between the index and outcome events. We use this process to create 16 distinct survival datasets, which we explain in detail in \cref{sec:dataDescription}. Each of these datasets corresponds to a different user behavior pattern in Aave and enables meaningful survival analyses. For each dataset, we used domain knowledge to derive 128 features describing the transaction and prior account history for the index event. To demonstrate the utility of our datasets, we define two benchmark tasks: (1) Time-to-Event Prediction, which involves estimating the expected time until an outcome event occurs, and (2) Event Occurrence Prediction, which involves predicting whether an outcome event will occur within a specified time frame. In addition to its scale and openness, our FinSurvival dataset fills a critical gap in the evaluation of survival models under real-world conditions of high-censoring and structured financial data. FinSurvival offers structured, high-dimensional data derived from real financial behavior with a mean censoring rate exceeding 80\% across its 16 datasets. Thus, this dataset provides the machine learning community with a much-needed testbed for developing and evaluating models that must perform reliably when event signals are rare, covariates are rich, and data is complex--conditions that are increasingly relevant in financial domains.

        \textbf{Contributions:} Our paper makes the following contributions:
        \begin{itemize}
            \item \textbf{Release of novel, large survival datasets}: We created and released a collection of 16 large-scale survival datasets derived from real financial transaction data. 
            \item \textbf{Benchmarking results:} We create two tasks for each of these survival datasets (time-to-event prediction and classification), then benchmark several models' performance for these tasks.
            \item \textbf{Open-source code:} All code written to reproduce the content of this paper is published in a Github repository~\footnote{\url{https://github.com/Large-Transaction-Models/DMLR_DeFi_Survival_Benchmark}}. This includes code to transform raw transaction data (along with a sample of raw transaction data) into survival data, code to compute data statistics in this paper, and code to reproduce the experiments.
        \end{itemize}

        The rest of this paper is organized as follows. In \cref{sec:dataDescription} we describe our datasets in detail and explain how they were converted into train and test sets for experiments. In \cref{sec:predictionTasks} we explain how we set up the survival prediction task for each dataset and provide benchmark results for several survival methods on this task. Similarly,  \cref{sec:classificationTasks} explains how we built a corresponding classification task using the Restricted Mean Survival Time (RMST)~[\cite{uno2014rmst}] for each dataset, once again benchmarking several classification methods on this task. Finally, we conclude with a discussion of the overall results in \cref{sec:discussion} and how this work can be extended in the future in \cref{sec:conclusionAndFutureWork}.

    \section{Dataset Description}
        \label{sec:dataDescription}
        
        \subsection{Overview of Survival Data}

\begin{figure}
                \centering
                \includegraphics[width=\linewidth]{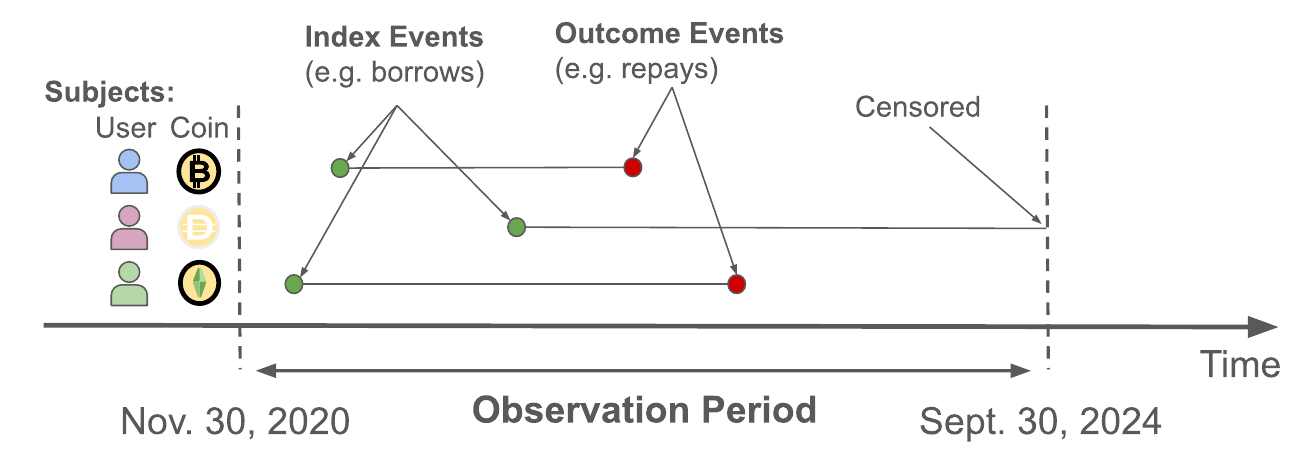}
                \label{fig:survDataPicture}
                \caption{The idea behind survival data. One or more subject types are selected and observed over a given observation period. Activity is monitored, waiting for an index event to trigger the start of a record. A chosen outcome event marks the end of the record, or the record is censored at the end of the observation period.}
            \end{figure}

            Survival analysis data are typically collected by identifying a cohort of subjects and recording the time until an event of interest, known as the ``outcome event,'' occurs. The starting point for measuring time is often referred to as the ``index event,'' which can be an initial diagnosis, treatment commencement, or any other significant starting point relevant to the study. Data collection involves tracking subjects over a specified period, noting whether and when the outcome event, such as death, relapse, or recovery, happens. Additionally, for those subjects who do not experience the outcome event within the observation period, their data are considered censored at the last point of follow-up, thereby accounting for incomplete observations. This method allows researchers to analyze the time-to-event data, accommodating both observed and censored cases, to derive meaningful insights into the factors influencing survival times. See \cref{fig:survDataPicture} for a visualization of this idea in the context of our data.

\begin{figure}
                \centering
                \includegraphics[width=\linewidth]{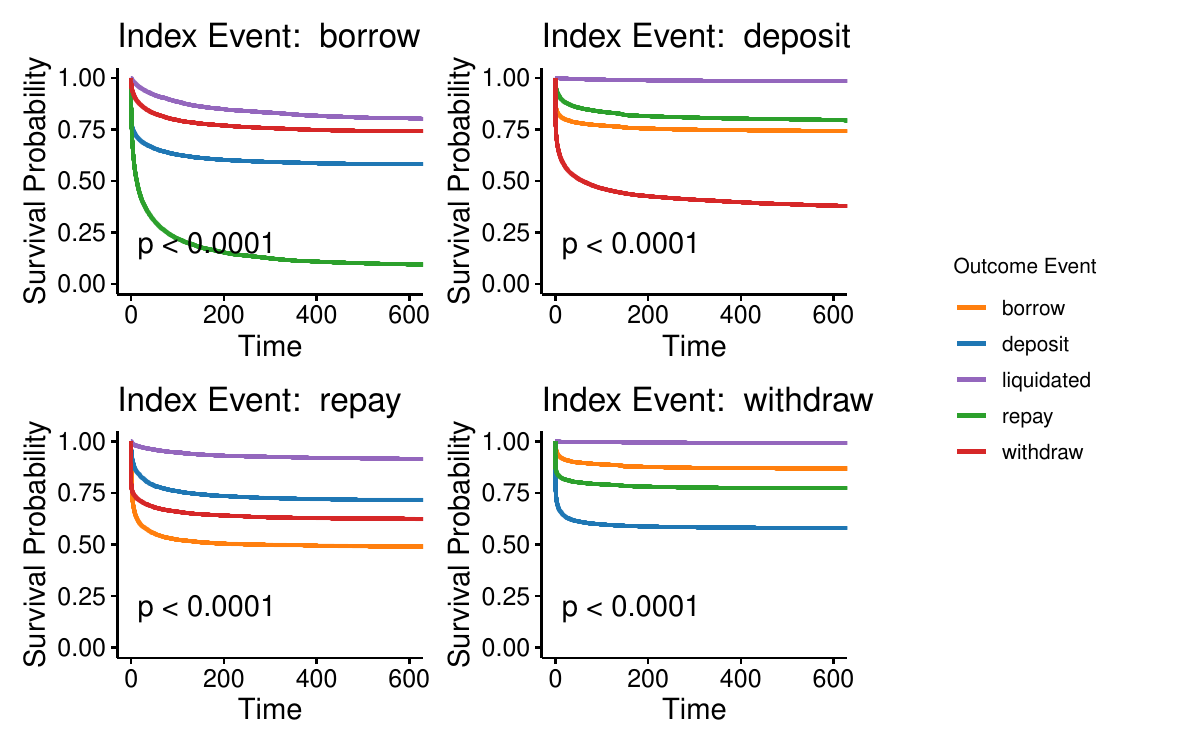}
                \label{fig:kmCurves}
                \caption{Kaplan--Meier curves for all index event and outcome event combinations. Each plot contains four curves representing a single index event and the four outcome events possible for that index event. These curves show that there is a variety of behaviors across our different datasets.}
            \end{figure}

            Our suite of survival datasets is all created from raw transaction data from the DeFi lending protocol Aave, specifically the Aave V2 Ethereum protocol. This data was acquired from The Graph\footnote{\url{thegraph.com}}. We built a pipeline to convert the raw transaction data into survival data, using this pipeline to create 16 survival datasets. These datasets were built by separately treating each transaction type (except for liquidations) as index events, and subsequently each of the other transaction types as a possible outcome event.  Liquidations, which roughly correspond to partial defaults on a borrowing transaction, are quite rare. This produces four possible index events, each with four possible outcome events. One notable feature of our data is that, because we were able to collect every transaction since Aave launched, we have no left-censored records. Every record has an associated index event. For an example of what one of our survival datasets looks like, see  \cref{tab:survDataStructure}. For an overview of all 16 datasets, see \cref{tab:allDatasetsOverview}. We also show Kaplan--Meier survival curves for all 16 datasets in \cref{fig:kmCurves} to show that these datasets represent different patterns of behavior.

            We also include 128 features for this survival data based on prior publications and domain knowledge [\cite{green2022defi,Green2023Clustering}]. The raw data for each transaction contains 22 features, most of which can be kept as features for each record. These include features like which coin was used in a transaction (reserve), how much money was involved in a transaction, the lending pool involved, etc. Different transaction types have slightly different inherent features, and if a feature is irrelevant for a particular transaction type, we leave it as NA. The exact features are listed in \cref{tab:rawTransactionFeatures} in \cref{sec:featureConstruction}.
            
            On top of these features, we engineered larger sets of features following three major themes. We built 19 features to represent the temporal aspect of transactions in different ways, encoding the date and time of each transaction into cyclic representations based on the day of the month, day of the week, day of the quarter, etc. We built 45 user-history features that represent up to the time of a given transaction a summary of the user's transaction history, such as how much money they have spent on each transaction type, how frequently they have made transactions, what coins they use most often, etc. Similarly, we built 40 features that represent a summarized history of the market as a whole up to the time of the transaction, such as how much of a specific coin in a transaction has been borrowed or deposited up to that point. A more detailed overview of these features can be found in \cref{sec:featureConstruction}.

            \begin{table}
                \caption{The structure of one of our survival datasets, Borrow-to-Repay, with some features excluded for brevity. ``Time'' represents the elapsed number of seconds between the index and outcome events. ``Status'' represents whether the observation was censored. ``User'' is a blockchain address of the user who performed the transaction. ``Coin'' is the cryptocurrency used in transactions. ``Index'' and ``Outcome'' are the index and outcome event types.}
                \rowcolors{1}{}{lightgray!25}
                \begin{tabularx}{\linewidth}{XXXXXXXX}
                    \hline
                    Time & Status & User & Coin & Index & Outcome & Amount (\$) & $\cdots$ \\
                    \hline
                    15,487 & 1 & 0xab123... & DAI & Borrow & Repay & 15,000.00 & $\cdots$ \\
                    190,601 & 1 & 0x98si4... & USDT & Borrow & Repay & 1,349.97 & $\cdots$ \\
                    $\vdots$ & $\vdots$ & $\vdots$ & $\vdots$ & $\vdots$ & $\vdots$ &$\vdots$& $\ddots$ \\
                    173,472 & 0 & 0x74flk... & USDC & Borrow & Repay & 598.80 & $\cdots$ \\
                    \hline
                \end{tabularx}
                \label{tab:survDataStructure}
            \end{table}

            \begin{table}[ht]
                \centering
                \caption{Overview of all the survival data in our suite of 16 datasets.}
                \rowcolors{1}{}{lightgray!25}
                \begin{tabularx}{\linewidth}{XX}
                     \hline
                     Attribute & Description \\
                     \hline
                     Total \# Records & 7,698,497 \\
                     Time Period & November 30, 2020 - September 30, 2024 \\
                     Subjects & Users, Coins \\
                     Unique Users & 114,861 \\
                     Unique Coins & 60 \\
                     Index Events & Borrow, Deposit, Repay, Withdraw \\
                     Outcome Events & Account Liquidated, Borrow, Deposit, Repay, Withdraw \\
                     Mean Censoring Rate & 81.26\%\\
                     \# Features & 128\\
                     Data Source & The Graph\\
                     \hline
                \end{tabularx}
                \label{tab:combinedDataSummary}
            \end{table}

            \begin{table*}[ht]
                \centering
                \caption{Summary statistics of the different types of survival data in the dataset based on index and outcome events.}
                \rowcolors{1}{}{lightgray!25}
                \begin{tabularx}{\linewidth}{XX|XXX|XX}
                    & & \multicolumn{3}{c}{FinSurvival Stats} & \multicolumn{2}{|c}{Classification Stats} \\
                    \hline
                    Index Event & Outcome Event & \# Records & Mean $~~$Delay &  Censored  \%& \# Records & Class 1 \% \\ 
                      \hline
                      borrow & liquidated & 264,536 & 323.30 & 83.42 & 255,209 & 4.77 \\ 
                      borrow & deposit & 640,497 & 246.85 & 87.50 & 251,730 & 16.83 \\ 
                      borrow & repay & 267,010 & 78.80 & 16.62 & 245,870 & 45.22 \\ 
                      borrow & withdraw & 578,605 & 274.43 & 89.50 & 252,816 & 11.55 \\ 
                      deposit & liquidated & 507,437 & 375.71 & 98.73 & 471,928 & 0.33 \\ 
                      deposit & borrow & 644,296 & 303.62 & 87.83 & 466,369 & 10.15 \\ 
                      deposit & repay & 595,761 & 313.39 & 86.43 & 468,721 & 10.16 \\ 
                      deposit & withdraw & 629,082 & 198.42 & 57.15 & 452,269 & 33.11 \\ 
                      repay & liquidated & 208,939 & 333.97 & 93.18 & 187,141 & 2.56 \\ 
                      repay & borrow & 226,764 & 192.42 & 57.65 & 177,426 & 34.35 \\ 
                      repay & deposit & 573,804 & 242.44 & 91.25 & 184,219 & 16.42 \\ 
                      repay & withdraw & 514,555 & 261.04 & 90.65 & 182,465 & 14.10 \\ 
                      withdraw & liquidated & 399,704 & 360.26 & 99.34 & 362,682 & 0.20 \\ 
                      withdraw & borrow & 555,420 & 298.85 & 93.26 & 358,695 & 6.76 \\ 
                      withdraw & deposit & 587,063 & 236.85 & 76.73 & 340,780 & 25.23 \\ 
                      withdraw & repay & 505,024 & 307.11 & 90.98 & 357,909 & 7.74\\ 
                      \hline
                \end{tabularx}
                \label{tab:allDatasetsOverview}
            \end{table*}

            \subsection{Train/Test Split}

                We had to split the data temporally into a training and a testing set to run experiments on this data. To do this, we chose a cutoff date of July 1, 2022. The training set includes data before this cutoff date, and the testing set includes data afterwards. This date was chosen because approximately 60\% of all transactions occur before this date and 40\% occur after. In addition to the cutoff, we included a buffer window at the end of both the training and testing sets, during which time we ignored new index events. We did this to give each index event a fair amount of time to see an outcome event before the end of the observation period. The buffer length was selected in conjunction with the restricted mean survival times (RMSTs) for each task. The RMST calculates the mean survival time of a dataset up to a specific point in time after the index event. When creating the classification task (see \cref{sec:classificationTasks}), the RMSTs were calculated for each task for increasing durations until the RMSTs changed by less than five percent. We used the converged RMST values for the classification criteria. Since these values ended up being less than 30 days (see \cref{tab:rmstVals}), we chose a 30-day buffer for ignoring new index events in the train and test sets.
                

    \section{FinSurvival Prediction Tasks and Results}
        \label{sec:predictionTasks}
        
            \subsection{Model Selection and Evaluation Metrics}

                We implemented four traditional survival regression models: Cox Proportional Hazards~[\cite{coxRegressionPaper}], Accelerated Failure Time (AFT)~[\cite{aftPaper}], Gradient Boosting Machine (GBM)~[\cite{gbmPaper}], and XGBoost~[\cite{xgboostPaper}]. The Cox model estimates hazard rates as a function of user and transaction features, while the AFT model predicts the log-transformed survival time under a Weibull distribution assumption. The GBM approach leverages boosting techniques to capture complex interactions. Finally, the XGBoost model applies an AFT objective to directly handle censored survival data. 

                We also implemented two deep-learning survival models: DeepSurv~[\cite{deepsurvPaper}] and DeepHit~[\cite{deephitPaper}]. Over the past decade, numerous deep survival models have been introduced, leveraging state-of-the-art deep learning architectures such as feed-forward neural networks and Transformer Models~[\cite{wiegrebe2024deep}]. Notably, there are Cox-based models like DeepSurv and discrete-time methods like DeepHit. DeepSurv uses a feed-forward neural network to model the log-risk function within a traditional Cox regression framework. In contrast, DeepHit treats time as discrete and typically employs classification techniques, with outcomes being binary event indicators for each discrete time point or interval. We implemented these two models using previously published code and compared their results with the traditional models on our data. 
                
                Detailed descriptions of each model’s implementation, including data preparation, feature selection, and hyperparameter settings, are provided in \cref{appendix:modelImplementations}.
                
                The Concordance Index (C-index) evaluated the baseline survival prediction models. The C-index assesses the models’ abilities to discriminate between individuals with different survival times. C-index values are better the closer they are to 1, and the lowest possible C-index is 0.5.

                To assess which models performed best overall, we computed the mean Borda rank for each model as described in \cite{pavaoRankingThesis}. First, we ranked the models on each dataset using the Borda ranking system, which assigns each candidate model a rank from $1$ to $n$, where $n$ is the number of candidates, based on their score. For example, since we have $n=6$ models, for any one dataset, the model with the highest c-index is given a rank of $1$, the second highest gets rank $2$, etc. With models being ranked for each dataset, we computed the mean Borda rank for each model to estimate which models were the best across the whole benchmark. We sort the columns according to this ranking. Lower Borda ranks mean better performance. 

                We use the same method to rank the difficulty of the datasets. By ranking how well individual models performed across all datasets and subsequently averaging these rankings, we can compare the difficulty of each dataset for our models to learn. We sort the rows in order of increasing mean Borda rank to represent the increasing difficulty of the datasets. Since both the rows and columns are sorted by their mean Borda rank, which is dependent on the number of rows or columns, we divide the mean Borda ranks by the number of rows and columns, respectively, to have the same scale on each axis.
                
        \subsection{FinSurvival Prediction Benchmark Results}

\begin{figure}
                        \centering
                        \includegraphics[width=\linewidth]{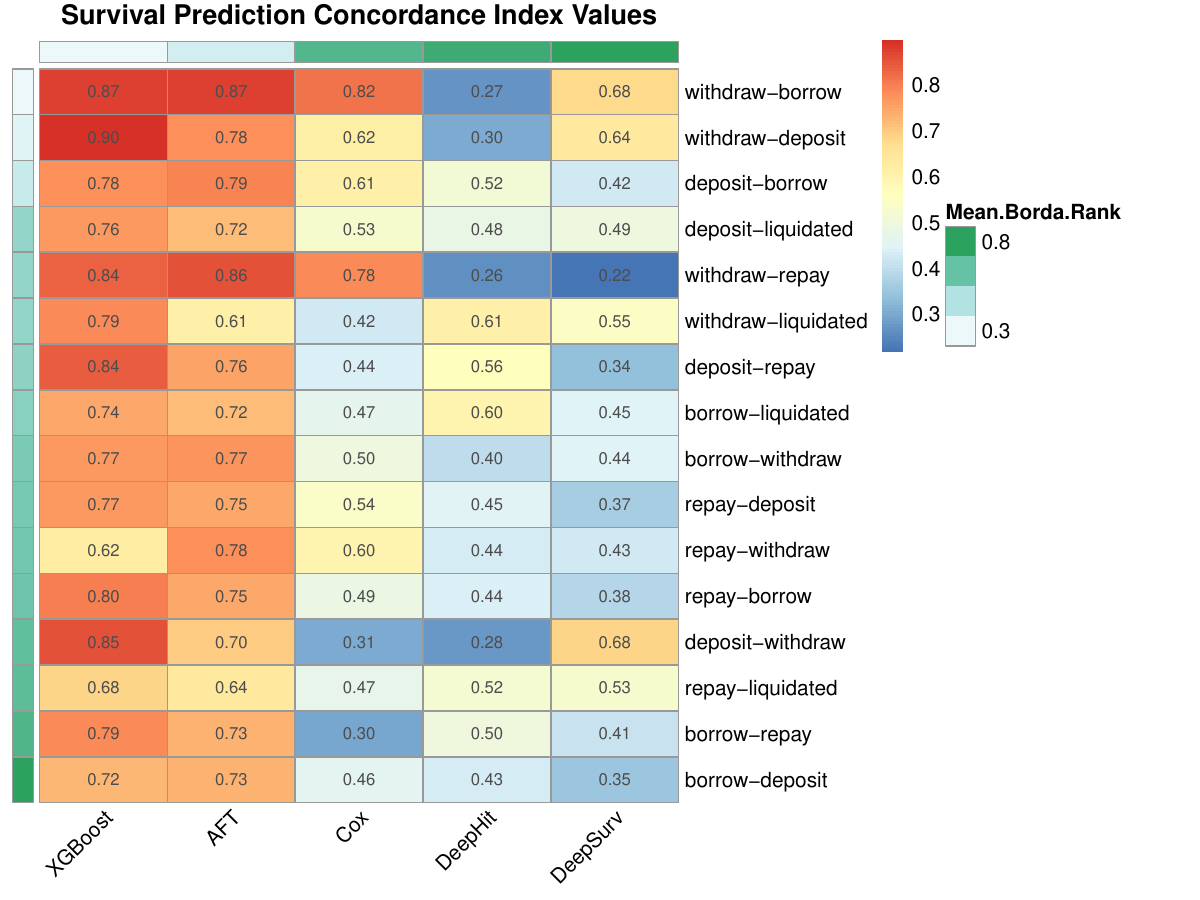}
                        \caption{Heatmap displaying prediction C-index values for survival outcomes across different index-outcome event pairs. Each cell represents one model's C-index score for an individual dataset. The rows and columns are each ordered in decreasing order based on the mean Borda rank among the rows and columns, respectively. Models on the left side performed better on average, and datasets towards the top of the heatmap were generally easier for models to learn.}
                        \label{fig:predictionResultsHeatmap}
                \end{figure}

                In \cref{fig:predictionResultsHeatmap}, we summarize the performance of the six survival models on our datasets (\cref{tab:predictionResults} in \cref{appendix:modelResults} gives the exact values). The rows and columns of the heatmap are sorted based on their mean Borda ranks.
                
                The XGBoost model consistently achieved the highest C-index values and had the highest average C-index overall. This was followed closely by AFT. While the Cox model ranked third overall, it really only performed well on two datasets (withdraw-borrow and withdraw-repay). The rest of the results for the Cox model amount to little more than random guessing. The GBM model performed extremely poorly, consistently attaining c-index values near or below 0.3. Since a c-index of 0.5 indicates random guessing, it would seem that GBM is consistently learning something about the data to predict survival times in a manner even worse than random guessing.

                The deep learning methods showed very poor performance. This highlights potential challenges in hyperparameter tuning and optimization and indicates the complexity inherent in applying deep learning to survival analysis tasks. The observed differences between linear and nonlinear models also emphasize the presence of complex nonlinear relationships within the data, underscoring the importance of careful feature selection, representation learning, and model choice.

                Overall, these results emphasize the utility of traditional survival analysis methods such as XGBoost and AFT for predicting financial events in DeFi environments. The wide variability in C-index values across different tasks shows this suite of datasets' inherent difficulty and complexity. These results also indicate that, while existing methods can be effective, there remains significant room for improvement in survival modeling, especially with deep-learning models.

                With the rows and columns arranged by mean Borda rankings, the heatmap shows the best-performing models on the left and the easiest-to-learn datasets on top. The models perform worse, and the datasets get harder as we move towards the bottom right. There are some interesting things to learn from this ranking. First, all four of the datasets that have withdrawal as the index event are ranked among the easiest six datasets. This feels a little bit counterintuitive since withdrawals involve pulling money out of the lending pool, and it does not feel like they set up an obvious next action for a user. Two of the more obviously meaningful datasets, deposit-withdraw and borrow-repay, are among the most difficult datasets to predict. These events could be easily interpreted as representing the time until a customer churns and the time until a loan repayment, respectively. The fact that they remain difficult to predict shows that user behavior is varied and based on a complex set of factors.  

    \section{FinSurvival-Classification Tasks and Results}
        \label{sec:classificationTasks}

        We converted the 16 survival tasks into a corresponding classification problem.  Converting survival analysis into classification tasks has gained traction, particularly as deep learning models redefine survival modeling by enabling binary classification within fixed time intervals~[\cite{wiegrebe2024deep}]. This approach allows neural networks and similar models to address the challenges of censored and high-dimensional data, which traditional survival methods, like the Cox Proportional Hazards model, often struggle with in complex datasets~[\cite{salerno2023highdimensional}]. Classification-based survival modeling has been applied effectively in fields such as finance and healthcare, where timely and interpretable predictions are critical~[\cite{deephitPaper}]. We now describe our strategy for picking the fixed time interval for each task. 

        \subsection{Classification data set creation} 

        \label{sec:classification}

            To structure survival data into a classification format, we used a Restricted Mean Survival Time (RMST) to determine a classification cutoff. RMST, which calculates the area under the survival curve up to a specific truncation time $\tau$, provides an interpretable measure of expected time-to-event within a fixed period. This method is particularly valuable for high-dimensional datasets, where traditional survival models, such as the Cox Proportional Hazards model, may struggle with censoring and dimensionality~[\cite{wiegrebe2024deep, salerno2023highdimensional}]. RMST was selected as it summarizes survival data without requiring the entire cohort to reach the event, offering a practical solution for heavily censored data.   Unlike median survival, which depends on at least 50\% of events occurring, RMST allows for summary statistics even with high censoring rates, making it a robust measure for our DeFi dataset. This approach aligns with the work of Uno et al~[\cite{uno2014rmst}], who recommend RMST for survival analysis tasks involving varying levels of censoring and complex event distributions~[\cite{uno2014rmst}].   RMST-based classification benchmark offers a standardized method for handling high-dimensional, censored survival data, crucial for evaluating model performance across complex financial datasets~[\cite{deephitPaper}].

             Table \ref{tab:rmstVals} shows the RMST value and the number of days $\tau$ used for each task. The RMST over $\tau$ days is used to convert survival times for each time to classes: class 1 if the survival time is $\le RMST$ and class 0 if the time to event is $>$ RMST.  Points that are censored in fewer than RMST days are dropped.  The RMST is calculated by measuring the area under the estimated Kaplan-Meier survival curve up to  $\tau$ days, so RMST($\tau$) monotonically increases with $\tau$. Thus, we want to pick a $\tau$ small enough to preserve data,  but also large enough so that the Kaplan-Meier curve has flattened out.  We calculated RMST($\tau$) for increasing values of $\tau$ in one-day increments and stopped when the change in RMST($\tau$) was less than 5\%. Class 1 is then defined as events that see their outcome in less time than RMST($\tau$). The RMST was calculated using the monotonic spline method, particularly useful for high-dimensional financial datasets, where complex survival patterns require precise modeling~[\cite{academic_biometrics2021}].
             \Cref{tab:allDatasetsOverview} shows the number of records and percentage of class 1 data in each FinSurvival Classification task. We note that some tasks are extremely unbalanced, with class 1 rates as low as 0.2\% 
             
            \begin{table}[htbp]
                \centering
                \caption{RMST Values and corresponding Days used to create FinSurvival-Classification}
                \label{tab:rmstVals}
                \rowcolors{1}{}{lightgray!25}
                \begin{tabularx}{\linewidth}{XXcc}
                    \hline
                    \textbf{Index Event} & \textbf{Outcome Event} & \textbf{Days} & \textbf{RMST}\\
                    \hline
                    Borrow  & Liquidated      & 20 & 17.43 \\
                    Borrow  & Deposit         & 21 & 19.88 \\
                    Borrow  & Repay           & 17 & 10.24 \\
                    Borrow  & Withdraw        & 21 & 20.21 \\
                    Deposit & Liquidated      & 20 & 17.52 \\
                    Deposit & Borrow          & 21 & 19.73 \\
                    Deposit & Repay           & 21 & 19.69 \\
                    Deposit & Withdraw        & 20 & 15.67 \\
                    Repay   & Liquidated      & 20 & 17.43 \\
                    Repay   & Borrow          & 20 & 14.99 \\
                    Repay   & Deposit         & 21 & 20.13 \\
                    Repay   & Withdraw        & 21 & 20.15 \\
                    Withdraw & Liquidated     & 20 & 17.72 \\
                    Withdraw & Borrow         & 21 & 20.26 \\
                    Withdraw & Deposit        & 21 & 18.14 \\
                    Withdraw & Repay          & 21 & 20.04 \\
                    \hline
                \end{tabularx}
            \end{table}

            \subsection{Model Selection and Evaluation Metrics}

                We evaluated a diverse set of popular classification methods to get a baseline.   
                The models used include Logistic Regression~[\cite{tibshirani1996regression}], Decision Tree~[\cite{breiman1984classification}],  XGBoost~[\cite{xgboostPaper}], and Elastic Net~[\cite{zou2005regularization}]. In addition to the four traditional models, we used two deep-learning models for classification: DeepHit~[\cite{deephitPaper}] and a neural net model of our implementation. See the appendix for details of packages used and how models were tuned. 
                
                To evaluate the performance of our classification models, we use the Area Under the Receiver Operating Characteristic Curve (AUC), which is a robust metric for assessing the discriminative capability of binary classification models even when datasets are imbalanced ~[\cite{aucScorePaper}]. The AUC score quantifies a model's ability to correctly rank examples by their predicted event occurrence probability, independent of classification thresholds. An AUC score closer to 1 indicates strong discriminative power, whereas an AUC closer to 0.5 suggests performance is no better than random guessing. As in the survival prediction task, we also aggregated the results from all 16 datasets to quantify which models were the best in each metric. We used the mean performance across all datasets for each model and the mean Borda rank for the models across each dataset.
                
                Due to some of the datasets being very imbalanced, we used the synthetic minority over-sampling technique (SMOTE~[\cite{chawla2002smote}]) on some of our datasets. For any dataset with 15\% or less of the data in Class 1 (see \Cref{tab:allDatasetsOverview}), if appropriate for the model being trained, we train two versions of the model: one with SMOTE applied to the training set and one without. We evaluate these two models' performance on a validation set taken from the training set and use the model with the better AUC for testing.

            \subsection{FinSurvival-Classification Benchmark Results}

\begin{figure}
                    \centering
                    \includegraphics[width=\linewidth]{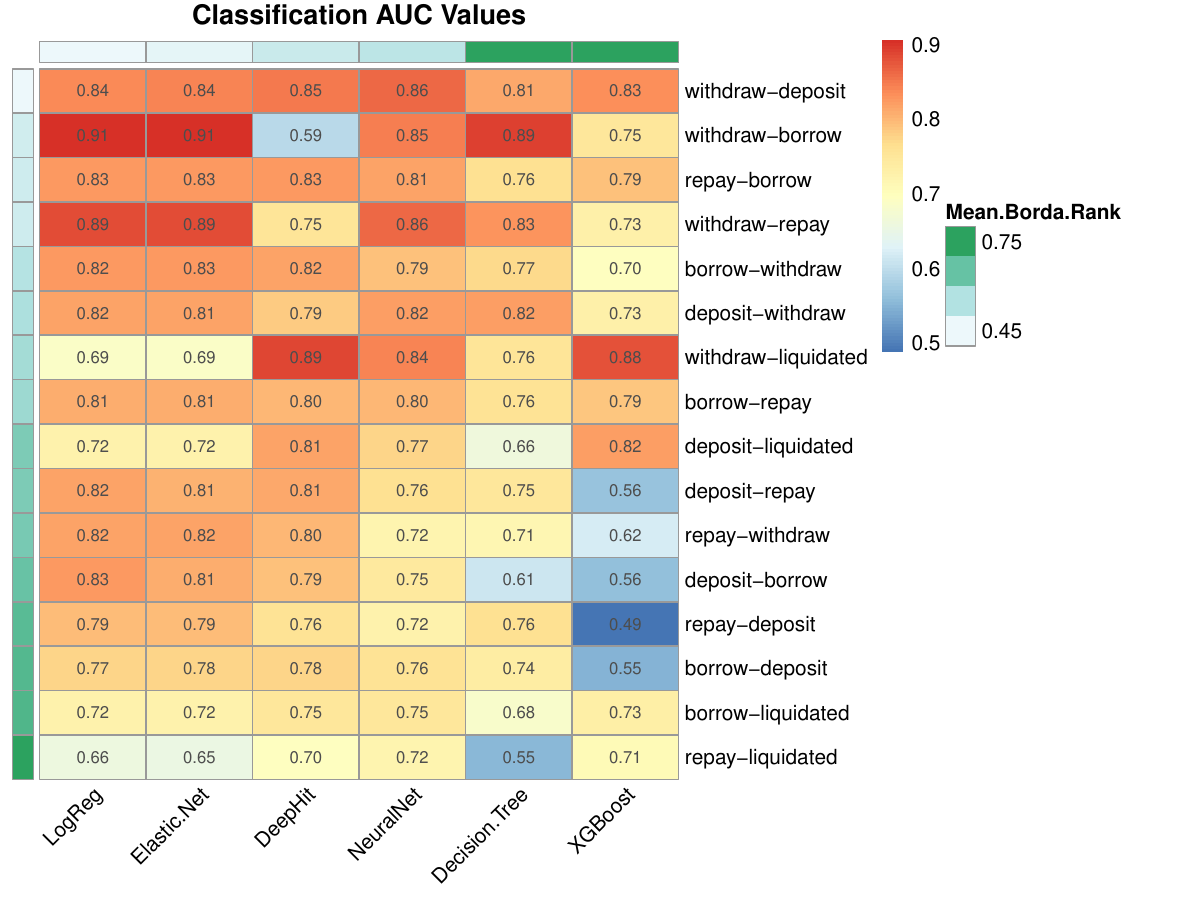}
                    \caption{Heatmap displaying classification AUC scores for survival outcomes across different index-outcome event pairs. Each cell represents one model's AUC score for an individual dataset. The rows and columns are each ordered in decreasing order based on the mean Borda rank among the rows and columns, respectively. Models on the left side performed better on average, and datasets towards the top of the heatmap were generally easier for models to learn.}
                    \label{fig:classificationResultsHeatmap}
                \end{figure}

                Figure~\ref{fig:classificationResultsHeatmap} presents a performance comparison of the six classification models across sixteen index-outcome event combinations (\cref{tab:classificationResults} in \cref{appendix:modelResults} gives the exact values). Each cell reports the AUC, capturing how effectively each model handles class imbalance and predictive precision.

                Surprisingly, the linear methods of logistic regression and elastic net performed the best. They performed almost identically, with logistic regression just edging out elastic net in mean AUC across all the tasks. The deep learning models of DeepHit and our custom neural net also performed strongly, achieving comparable results to the linear models. In contrast to the results on the prediction task, XGBoost performed very poorly compared to the other models.

                Overall, these classification results show that linear models perform well on this task for datasets and demonstrate the importance of model selection for different tasks. The deep-learning models show improved results compared to deep learning models on the prediction task, suggesting that deep learning models are easier to train for classification tasks versus survival prediction tasks. The collective results show that there are a variety of difficulties across these tasks as well.

                As with the prediction results, the rows and columns are arranged according to the mean Borda rank of each respective axis. Similar to the prediction results, the two easiest datasets are withdraw-borrow and withdraw-deposit (although in the opposite order). For this task, though, deposit-withdraw and borrow-repay are significantly easier than they were for the prediction task, ranking in the top half in terms of difficulty. This is a significant jump, and interesting in that it suggests that while it can be very difficult to predict how long it will take for users to churn or repay loans, it can be much more feasible to threshold the duration and predict whether the event will happen before the threshold.

    \section{Discussion}
        \label{sec:discussion}
        The FinSurvival benchmark underscores the challenges of modeling high-dimensional, heavily censored DeFi data. In survival prediction, traditional models such as XGBoost and AFT regression outperformed deep learning methods, with XGBoost delivering the highest discriminative power. XGBoost and AFT actually performed fairly well, and the deep learning methods exhibited quite poor performance. This could be due to poor tuning, but we spent significant time tuning these methods, trying to achieve comparable results, and were not able to do so. These results suggest that conventional approaches remain solid in complex financial settings despite recent deep learning advances, and that there is still a lot of progress to be made in deep-learning models for survival analysis.

        For classification tasks using RMST-based thresholds, basic logistic regression and Elastic Net led in AUC score, and deep-learning methods demonstrated moderate performance. However, tree-based methods like XGBoost exhibited mixed or poor results. Although the deep learning models tested show promise, their performance may have been constrained by limited tuning. Further optimization could potentially improve their ability to capture the nuances of the dataset.

        Our results demonstrate that the FinSurvival dataset is demanding and complex, offering a rigorous benchmark for evaluating and advancing AI survival models. This work highlights opportunities for improved deep learning strategies and hybrid approaches to better capture the intricate patterns inherent in large-scale financial survival data.
        
            
        \impact{Our FinSurvival benchmark and datasets fill a critical need for large-scale survival datasets in finance. Unlike traditional survival data limited to small samples or sensitive domains like medical records, our openly accessible DeFi-based transaction data offers millions of records covering loans, deposits, and liquidations. By capturing high censoring rates and complex user behaviors at scale, FinSurvival drives methodological innovation in deep learning for survival modeling. It can also be helpful for quantifying things like credit risks, repayment patterns, etc., in a transparent manner. The openness and scale of this benchmark fill in an important gap that should help improve the accuracy and modeling capabilities of AI methods for survival analysis and help extend survival research to new domains. While there are potential ethical, privacy, and fairness concerns associated with the release and use of traditional financial data, these are largely ameliorated in these DeFi sets. DeFi users inherently consent to make their account transactions public by engaging with a DeFi protocol on a public blockchain. There is no personal or private information associated with account IDs. Also, DeFi protocols such as Aave are inherently fair because every action of the protocol is encoded in published smart contracts that apply identically to all accounts as described by public data on the blockchain. This contrasts with the well-documented biases that have been found in traditional lending, e.g., red-lining in mortgage lending \cite{yinger2018discrimination}. The possibility that the survival models produced could result in unethical and fraudulent behavior is very small.}

    \section{Conclusion and Future Work}
        \label{sec:conclusionAndFutureWork}

        This suite of survival datasets represents only a small fraction of the transaction data available through DeFi protocols. Not only are there numerous other deployments of the Aave protocol on other blockchains that have far more transaction data than the Ethereum chain used for these datasets, but there are other lending protocols, and other DeFi protocol types like Decentralized Exchanges (DEXs) that generate huge amounts of transaction data for unique problems. Our data creation pipeline can easily be used to convert transaction data from all of these protocols into new, massive survival data to create additional use cases and further challenges for survival modeling. 

        One way these datasets could be expanded is by adding more features. An interesting avenue of expansion could be incorporating exogenous data such as stock prices and cryptocurrency prices over time, along with other data typically associated with price data, like trade volume, volatility, etc. These could be joined with our existing data based on the timestamp and may help models have more predictive power.

        The features created for this work were hand-crafted based on domain knowledge and intuition as to what features might be useful for prediction. This process takes a long time and extensive domain knowledge. One way this data could be used better is through an automatic feature-creation process that leverages AI techniques such as BERT-style transformer-based models to auto-generate more interesting features based on the raw transaction data. Such a model could create embeddings for each index event that encode a user's entire transaction history and capture more mathematically complex relationships in the data.

        One limitation of this work, as it currently stands, is that it does not take into account competing risks. Different outcome events can ``compete'' with one another as possible outcomes for the same index event. For instance, if a user's borrowed funds are liquidated, this competes with their ability to repay that loan, i.e., borrow. Our survival modeling pipeline can include competing events in creating datasets, but we have not included them in this analysis~[\cite{green2024defi}]. Incorporating competing events in these analyses would help the models more accurately reflect the dynamics of user behavior within the underlying transaction data.


    \acks{
        We would like to thank students in the Data Analytics Research Course at Rensselaer who contributed to the creation of this project: Kaiyang Chang, Sean Fitch, Haolin Luo, Charlotte Newman, Emmet Whitehead, Daniel Kirtman, Campbell Drahus, Alejandro Laphond, and David Quintero.
        The authors acknowledge the support from NSF IUCRC CRAFT center research grants (CRAFT Grant \#22015) for this research. The opinions expressed in this publication and its accompanying code base do not necessarily represent the views of NSF IUCRC CRAFT.
        }
    
    \bibliography{references}
    
    \appendix

         \section{Model Implementation Details}
            \label{appendix:modelImplementations}

            For both the survival prediction and classification tasks, our default data preprocessing involves three main steps: (1) scaling the data, (2) encoding categorical data, and (3) applying Principal Component Analysis (PCA).

            For scaling the data, we mean-center the training data using R's built-in \verb|scale| function. We use the training mean and standard deviation to scale the testing data as well. This scaling is only applied to numeric features.

            We do two things to handle categorical data. First, for any categorical feature with more than ten categories, we keep the ten most frequent categories from the training set and combine the remaining into another category called ``Other''. We keep the same ten categories in the testing set and make ``Other'' for any remaining data as well. We do this so models have an easier time learning about less-common categories and to ensure the categories seen in the testing set are the same as in the training set. After this, we apply one-hot encoding to the categorical features using the \verb|fastDummies| package in R~[\cite{fastDummiesPackage}].

            Finally, we apply PCA to the resulting data to reduce collinearity among the features. We use the \verb|prcomp| function from \verb|stats| package~[\cite{statsPackage}]. We compute the principal components on the training set and keep the principal components that explain 90\% of the variance in the data. We use the same set of principal components on the testing set.

            Unless otherwise noted, all the subsequent models for prediction and classification tasks follow the above steps in their data preprocessing.
        
            \subsection{Survival Prediction Models}
                
                \textbf{Cox Proportional Hazards Model:} The Cox Proportional Hazards regression model is used to relate various factors, considered simultaneously, to survival time or time-to-event. In this model, the measure of effect is the hazard rate, which is the probability that the event of interest occurs, given that the participant (data point) has survived up to a given time. Our Cox proportional hazards model implementation is trained using the \verb|coxph| function from the survival package~[\cite{survival-package}]. We use the default parameters to train the model.

                \textbf{Accelerated Failure Time Model:} Accelerated Failure Time models~[\cite{aftPaper}] are an alternative to the commonly used proportional hazards model. In an AFT model, the effect of the covariates accelerates or decelerates the survival time by a specific factor. This acceleration factor is assumed to be constant. Our implementation of the AFT model uses the \verb|survreg| function from the \verb|survival| package, and we use the Weibull distribution to fit the data. Any other parameters use the default values.
  
                \textbf{XGBoost:} Extreme Gradient Boosting (XGBoost~[\cite{xgboostPaper}]) is a distributed gradient-boosting decision tree algorithm for regression and classification. Our implementation of XGBoost does not use PCA in its data preprocessing. We train the model using the \verb|xgb.train| function from the \verb|xgboost| package~[\cite{xgboostPackage}]. We use negative log-likelihood as our loss function. We use a maximum of 1,000 boosting iterations and use a validation set to check the loss every 50 iterations, stopping early if the validation loss does not improve.

                \textbf{DeepSurv:} The DeepSurv model was published by \cite{deepsurvPaper} and fits a neural network based on the partial likelihood from a Cox Proportional-Hazards model. Our implementation uses the \verb|deepsurv| function from the \verb|survivalmodels| package~[\cite{survivalmodelsPackage}]. After experimenting with various hyperparameter configurations based on experiments from the original paper, we followed its configuration from the ``simulated linear experiment.'' We had difficulties training this model on our full datasets, so we used uniform random sampling to sample 40,000 rows from each training set and 20,000 rows from each testing set to fit and evaluate the model.

                \textbf{DeepHit:} The DeepHit model was published by \cite{deephitPaper} and fits a neural net based on the probability mass function of a discrete Cox model. Our implementation uses the \verb|deepHit| function from the \verb|survivalmodels| package~[\cite{survivalmodelsPackage}]. To train DeepHit, we used the following hyperparameter configuration (any omitted parameters use the package defaults):

                \begin{table}[hbt!]
                    \centering
                    \caption{Hyperparameters used for training DeepHit.}
                    \rowcolors{1}{}{lightgray!25}
                    \begin{tabularx}{\linewidth}{XX}
                        \hline
                        Parameter Name & Value \\
                        \hline
                        epochs & 10 \\
                        cuts & 50 \\
                        optimizer & sgd \\
                        l2\_reg & 4.425 \\
                        lr\_decay & 3.173e-4 \\
                        momentum & 0.936 \\
                        patience & 0 \\
                        lr & 0.001 \\
                        \hline
                    \end{tabularx}
                    \label{tab:deepHitHyperparameters}
                \end{table}

        \subsection{Classification Models}

            All the classification models follow the same default data preprocessing as the prediction models unless otherwise noted. After the data has been transformed into the binary classification task, unless stated otherwise, the models use the following policy for applying Synthetic Minority Oversampling Technique (SMOTE)~[\cite{chawla2002smote}]: If the training data has less than 15\% of its data in class 1 (see \cref{tab:allDatasetsOverview} for class 1 \%s), then a validation set is created from the training data and two models are fit to the training data without the validation set: one model on the raw training data and one model on the training data with SMOTE applied. These models are then tested on the validation set to see whether SMOTE helped. The better of the two models is used to predict on the testing set. All combinations of model type and dataset for which SMOTE was used are indicated in \cref{tab:classificationResults} with an asterisk (*).
            
            \textbf{Logistic Regression (LogReg)}:
                Logistic regression is a simple yet effective linear model that estimates the probability that a given input belongs to a particular class. It works by fitting a weighted sum of the input features to a sigmoid curve, which outputs values between 0 and 1 representing predicted probabilities. We use the \verb|glm| function from the \verb|stats| package~[\cite{statsPackage}] for logistic regression.

            \textbf{Decision Tree}: 
                A decision tree splits the data into smaller and smaller subsets based on simple feature-based rules, eventually making predictions at the leaves of the tree. It is easy to interpret and can capture non-linear relationships, but may overfit if not pruned or regularized. We used the \verb|rpart| function from the \verb|rpart| package~[\cite{rpartPackage}] for our decision tree implementation. We used a max depth of 30 and a minimum split of 20.
                
            \textbf{XGBoost}: 
                We used the same implementation of XGBoost as described above in the prediction models, but tuned it slightly differently. For the classification task, we use \verb|logloss| as the evaluation metric. Our max depth is set to 6. We use eta=0.1, subsample=0.8, colsample\_bytree=0.8, and scale\_pos\_weight is based on the relative proportion of class 0 data to class 1 data in the training set. We train with 200 rounds and early stopping rounds set to 10.
            
            \textbf{Elastic Net}:
                Elastic Net is a linear model that combines both Lasso (L1) and Ridge (L2) regularizations. It is useful when there are many correlated features, as it tends to select groups of them while also shrinking their magnitudes to avoid overfitting. We used the \verb|glmnet| function from the \verb|glmnet| package for our implementation~[\cite{glmnetPaper, glmnetPaper2}].
    
            \textbf{DeepHit}:
                Our implementation of DeepHit for the classification problem uses Python and the \verb|reticulate| package in R~[\cite{reticulatePackage}]. This implementation is based on the code from the original paper, using the model for a binary classification task. We train on 10 epochs with a learning rate of 0.001.

            \textbf{Neural Net}:
                Neural networks consist of multiple layers of interconnected nodes, or neurons, that apply non-linear transformations to the input data. They are highly flexible and can capture complex patterns, but require careful tuning and lots of data to avoid overfitting. We built our own relatively simple neural net architecture for this classification task. The model uses a feed-forward architecture implemented with Keras' Sequential API. It consists of an input layer accepting features of a specified dimension, two hidden layers, each with 64 units and ReLU activation, followed by batch normalization and a dropout layer with a 0.2 dropout rate. Then there's a custom transformation layer that applies a $\log(1+x)$ operation to simulate a transformation relevant to risk modeling. Finally, there's an output layer with two neurons and softmax activation for binary classification. The model is compiled with the Adam optimizer, categorical cross-entropy loss, and gradient clipping to mitigate exploding gradients.

    \subsection{Model Results}
        \label{appendix:modelResults}
        \Cref{tab:predictionResults} and \cref{tab:classificationResults} give the full numeric results for all models across all datasets.
        \begin{table*}[htb!]
            \centering
            \rowcolors{1}{}{lightgray!25}
            \caption{Performance evaluation of survival regression models on various datasets using the Concordance Index. Best results per row are in bold.}
            \label{tab:predictionResults}
            \begin{tabularx}{\textwidth}{X|XXXXXX}
                
                Dataset & XGBoost & Cox & AFT & DeepHit & DeepSurv \\ 
                \hline
                borrow-repay & \textbf{0.788} & 0.295 & 0.734 & 0.498 & 0.412 \\ 
                borrow-deposit & 0.723 & 0.459 & \textbf{0.730} & 0.434 & 0.349 \\ 
                borrow-withdraw & 0.769 & 0.496 & \textbf{0.772} & 0.400 & 0.443 \\ 
                borrow-liquidated & \textbf{0.745} & 0.469 & 0.721 & 0.596 & 0.449 \\ 
                repay-borrow & \textbf{0.798} & 0.489 & 0.745 & 0.442 & 0.383 \\ 
                repay-deposit & \textbf{0.768} & 0.544 & 0.749 & 0.455 & 0.366 \\ 
                repay-withdraw & 0.620 & 0.599 & \textbf{0.778} & 0.436 & 0.426 \\ 
                repay-liquidated & \textbf{0.681} & 0.474 & 0.640 & 0.519 & 0.528 \\ 
                deposit-borrow & 0.782 & 0.609 & \textbf{0.793} & 0.515 & 0.423 \\ 
                deposit-repay & \textbf{0.840} & 0.438 & 0.756 & 0.561 & 0.337 \\ 
                deposit-withdraw & \textbf{0.854} & 0.305 & 0.698 & 0.277 & 0.683 \\ 
                deposit-liquidated & \textbf{0.763} & 0.526 & 0.722 & 0.479 & 0.492 \\ 
                withdraw-borrow & \textbf{0.874} & 0.816 & \textbf{0.874} & 0.269 & 0.675 \\ 
                withdraw-repay & 0.835 & 0.785 & \textbf{0.856} & 0.261 & 0.218 \\ 
                withdraw-deposit & \textbf{0.899} & 0.620 & 0.779 & 0.303 & 0.640 \\ 
                withdraw-liquidated & \textbf{0.788} & 0.423 & 0.608 & 0.608 & 0.551 \\ 
                \hline
                \hline
                Mean & \textbf{0.783} & 0.522 & 0.747 & 0.441 & 0.461 \\ 
                Mean Borda Rank & \textbf{1.312} & 3.625 & 1.750 & 4.000 & 4.312 \\ 
                \hline
            \end{tabularx}
        \end{table*}

        \begin{table*}[htb!]
            \centering
            \rowcolors{1}{}{lightgray!25}
            \caption{Performance evaluation of classification models on various datasets using AUC Scores. Best results per row are in bold. Entries with * used SMOTE.}
            \label{tab:classificationResults}
            \begin{tabularx}{\linewidth}{X|XXXXXX}
                Dataset & LogReg & Decision Tree & XGBoost & Elastic Net & DeepHit & NeuralNet \\ 
                \hline
                borrow-liquidated & 0.723 & 0.681 & 0.732 & 0.723 & \textbf{0.750} & 0.748 \\ 
                borrow-deposit & 0.774 & 0.737 & 0.549 & \textbf{0.775} & \textbf{0.775} & 0.758 \\ 
                borrow-repay & \textbf{0.809} & 0.757 & 0.789 & \textbf{0.809} & 0.799 & 0.801 \\ 
                borrow-withdraw & 0.825 & 0.769 & 0.697 & \textbf{0.826} & 0.817 & 0.793 \\ 
                deposit-liquidated & 0.725* & 0.661 & \textbf{0.823} & 0.723* & 0.815 & 0.774 \\ 
                deposit-borrow & \textbf{0.827} & 0.610 & 0.561 & 0.810 & 0.793 & 0.747 \\ 
                deposit-repay & \textbf{0.816} & 0.749 & 0.564 & 0.806 & 0.813 & 0.761 \\ 
                deposit-withdraw & 0.816 & 0.821 & 0.731 & 0.815 & 0.785 & \textbf{0.820} \\ 
                repay-liquidated & 0.659 & 0.552 & 0.708 & 0.652 & 0.697 &\textbf{0.720} \\ 
                repay-borrow & \textbf{0.827} & 0.762 & 0.792 & \textbf{0.827} & 0.826 & 0.815 \\ 
                repay-deposit & \textbf{0.794} & 0.764 & 0.488 & \textbf{0.794} & 0.759 & 0.723 \\ 
                repay-withdraw & 0.818 & 0.715* & 0.618 & \textbf{0.819} & 0.800 & 0.721 \\ 
                withdraw-liquidated & 0.687* & 0.756* & 0.880 & 0.687* & \textbf{0.890} & 0.841 \\ 
                withdraw-borrow & \textbf{0.907} & 0.892 & 0.749 & \textbf{0.907} & 0.594 & 0.847 \\ 
                withdraw-repay & \textbf{0.886} & 0.829 & 0.731 & \textbf{0.886} & 0.752 & 0.863 \\ 
                withdraw-deposit & 0.837 & 0.811 & 0.834 & 0.840 & 0.852 & \textbf{0.862} \\ 
                \hline
                \hline
                Mean & \textbf{0.795} &  0.741 & 0.702 &  0.793 & 0.782 & 0.787 \\
                Mean Borda Rank & \textbf{2.59} & 4.75 & 4.75 & 2.69 & 3.03 & 3.19  \\
                \hline
            \end{tabularx}
        \end{table*}

        \section{Survival Data Creation Pipeline}
            \label{sec:dataCreationPipeline}
            \subsection{Raw Transaction Data Collection}

                The first phase for creating this dataset was collecting the raw market transaction data from the Aave platform. The underlying transaction data upon which this data is based comes from The Graph\footnote{thegraph.com}, a decentralized protocol for indexing and querying data from blockchains that primarily target the Ethereum network. Specifically, we collect transaction data from the Aave V2 subgraph\footnote{https://thegraph.com/hosted-service/subgraph/aave/protocol-v2?version=current}. This subgraph contains many tables necessary to query to get a comprehensive view of the transactions in Aave. 
    
                The data for each transaction type is in its table, so we queried each type to get data for all transactions from November 30, 2020, through September 30, 2024. The transaction types we use in our dataset are Deposits, Withdrawals, Loans, Repayments, and Liquidations. 
    
                To get all pertinent information for each transaction, we also collect data from the ReserveParamsHistoryItems from the same time span. It provides timestamped information about reserves whenever they are used in a transaction within Aave. This is how we can get information about how much a reserve was worth and what its interest rates were at the time of each transaction. 
    
                Finally, we also collected data about the individual coins (the table is called Reserves) to get basic information about each reserve, such as its symbol, what functionality is available for it in Aave, and how many decimal places to adjust its numerical values by. With the information from all of these tables, we were able to create one unified and human-readable view of the transaction data. We combined all of the transaction-type-specific tables, sorted them chronologically by their UNIX timestamp, and replaced IDs in many columns with more pertinent information so that each transaction could be comprehensible to a human. Information about the coins used in each transaction was added explicitly to each transaction record, so that at each transaction we can see the symbol of the coin(s) involved (e.g., BTC, ETH). The amounts of each currency being used in each transaction were adjusted by the currencies' specific decimal exponent, as well as their conversion factors to USD at each transaction time. Other time-dependent, currency-specific data was added to each transaction to increase the amount of information contained within one record.
                
                The final structure of this transaction-level data is shown in \cref{tab:rawTransactionStructure}. This table does not include all the columns of the data, as there are too many features to include and many of them are transaction-type-specific. We also provide a table of metadata about these transactions in \cref{tab:transactionMetadata}.

                \begin{table*}[ht]
                    \centering
                    \caption{Structure of raw transaction data showcasing the main features present for all transaction types.}
                \rowcolors{1}{}{lightgray!25}
                    \begin{tabularx}{\textwidth}{p{.15\textwidth}XXXXXp{0.025\textwidth}}
                        \hline
                         Datetime & Type & User & Coin & Amount & Amount (\$) & $\cdots$ \\
                        \hline
                        11-30-2020 23:15:00 & Deposit & $<$ID$>$ & USDT & 100.00 & 100.00 & $\cdots$ \\
                        11-30-2020 23:15:30 & Borrow & $<$ID$>$ & XSUSHI & 15.52 & 100.00 & $\cdots$ \\
                        $\vdots$ & $\vdots$ & $\vdots$ & $\vdots$ & $\vdots$ & $\vdots$ & $\ddots$ \\
                        12-31-2023 23:50:00 & Repay & $<$ID$>$ & DAI & 25,000.667 & 24,978.34 & $\cdots$ \\
                        12-31-2023 23:50:45 & Withdraw & $<$ID$>$ & WETH & 3.652 & 8,976.09 & $\cdots$ \\
                        \hline
                    \end{tabularx}
                    \label{tab:rawTransactionStructure}
                \end{table*}

            \begin{table}[ht]
                \centering
                \caption{Metadata about the transaction data that we collected, cleaned, and used to create the survival data being published with this paper.}
                \rowcolors{1}{}{lightgray!25}
                \begin{tabularx}{\linewidth}{XX}
                    \hline
                    Attribute & Description \\
                    \hline
                    Total Transactions & 1,977,491 \\
                    Total Users & 117,008 \\
                    Number of Features & 38 \\
                    Time Span & November 30, 2020 - September 30, 2024 \\
                    Size of CSV & 668.8 MB \\
                    Data Source & The Graph \\
                    \hline
                \end{tabularx}     
                \label{tab:transactionMetadata}
            \end{table}
            
        \subsection{Transformation to Survival Data}
        
             Phase two of creating this data involved the creation of the pipeline to convert transaction data into survival data. Survival data, also known as time-to-event data, involves observations where the outcome of interest is the time until a specific event occurs. This type of data is characterized by two main components: the observed time \( T_i \), which is either the time until the event occurs or the time until the last follow-up (for censored data), and the event indicator \( \delta_i \), which denotes whether the event has occurred (\(\delta_i = 1\)) or the observation is censored (\(\delta_i = 0\)). Additionally, survival data often includes a set of covariates \( \mathbf{X}_i \) that represent other variables which might influence the time-to-event. Mathematically, a survival dataset for \( n \) subjects is represented as \(\{(T_i, \delta_i, \mathbf{X}_i) : i = 1, 2, \ldots, n\}\). This structure enables the analysis of both the timing of events and the factors that affect these timings.

            The motivating idea for how our DeFi transaction data can become survival data is that each transaction a user performs could be considered an ``index event,'' triggering the start of a record which lasts until a future transaction of interest, which could be considered the ``outcome event.'' For instance, if we want to track how long it takes for a user to pay off a loan of a certain currency, we could treat the transaction where a user borrows that currency as the index event and track that user until they make a repayment of that same currency. With this idea in mind, our pipeline for creating survival data needs the following parameters: 

            \begin{itemize}
                \item \textbf{Event Data:} A tabular dataset containing all recorded events that could be relevant to creating the survival dataset.
                \item \textbf{Subjects:} A specified set of one or more columns of the event data that define who/what the subjects will be that we track when creating the survival data. For instance, most of the time we want to see both the ``user'' and the ``coin'' columns, because we are interested in how an individual user interacts with a specific coin over time.
                \item \textbf{Observation Period:} A start and end date and time over which to compute the survival data. By default, this can be the entire duration of the events dataset, but could be set to a shorter time window for more targeted analysis.
                \item \textbf{Index Event Set:} One or more event types which will be treated as the index events to start the tracking of survival records.
                \item \textbf{Outcome Event Set:} One or more event types that will trigger the end of a survival record if it occurs following an index event made by the same subject(s).
            \end{itemize}

            With these parameters defined, we create the pipeline for survival data creation. Given all the parameters, we first filter out all events that do not occur within the specified observation period. Then we create a subset of the data that only includes the desired index events and another subset that only includes the desired outcome events. Grouping these subsets by the selected subjects, we then perform a rolling join on the index events with the outcome events, matching the subjects and using the first event after each index event. This creates a table where each row has information on an index event and the first outcome event performed by the same subject, if any, performed after the index event. With this, we can calculate the elapsed time between the two events, using the final time of the observation period as the outcome event time if no appropriate outcome event occurred.
            
        \subsection{Curating Survival Datasets}
        
            Considering each of the five transaction types as possibilities for index and outcome events was the obvious way to go about this, but liquidation events need to be handled more carefully than the others due to their involving multiple parties. So, first we considered the ``basic'' transactions of borrows, repays, deposits, and withdraws. If we choose one of these transaction types as an index event (e.g., borrows) and another type as outcome events (e.g., repays), we can create survival data that answers a question like ``How long do users take to repay after borrowing?'' It is important to note that the choice of subjects here is both the user and the currency involved in the transaction, because if an index event shows a user borrowing e.g. Wrapped Bitcoin, an appropriate outcome event should be that same user repaying Bitcoin, not a different currency for which they might also have a loan. Putting these ideas together, we created 12/16 datasets using the different combinations of basic transactions as the index and outcome events.

            We wanted to include survival datasets using liquidations. Liquidations can occur when a user's overall account in Aave has an ``unhealthy'' balance of deposited assets compared to borrowed assets. These assets can include a variety of different currencies. When a user's account is unhealthy, another user can perform a liquidation transaction, paying off a portion of the unhealthy user's loans to claim an equivalent portion of the user's deposited collateral assets and a small liquidation bonus from the protocol as an incentive. So, liquidation transactions include more than just one user. They include a ``liquidator'', which is the user who performs the liquidation transaction, and a ``liquidatee'' whose account is being liquidated. Additionally, a liquidation transaction can involve any one of a user's borrowed currencies as the ``principal'' currency, and any one of the user's deposited currencies as the ``collateral'' currency. Given all of this, we handle liquidations differently than the other transaction types. In our suite of datasets, we only consider the case when a user's account is liquidated. We use this event exclusively as an outcome event.

        \subsection{Feature Construction}
            \label{sec:featureConstruction}
            The feature engineering process produced a total of 128 features derived from raw transaction data on AAVE Mainnet V2, of which 106 are constructed from the base features. These features fall into four primary categories: base features, user history features, market history features, and time features. The base features refer to the original fields extracted from the raw data, such as transaction amounts, timestamps, and coin types. From these base features, additional derived features were created to capture more complex relationships and non-linear patterns relevant to survival prediction up to and including the index event. The full list of base features can be found in \cref{tab:rawTransactionFeatures}.

            \begin{table}[ht]
                \centering
                \caption{The 24 features from the raw transaction data and the transaction types for which each feature is relevant.}
                \rowcolors{1}{}{lightgray!25}
                \begin{tabularx}{\linewidth}{XX}
                  \hline
                  Feature Name & Relevant Transaction Types \\ 
                  \hline
                  timestamp & All \\ 
                  user & All \\ 
                  pool & All \\ 
                  type & All \\
                  reserve & Borrow, Deposit, Repay, Withdraw \\ 
                  coinType & Borrow, Deposit, Repay, Withdraw \\
                  amount & Borrow, Deposit, Repay, Withdraw \\ 
                  amountUSD & Borrow, Deposit, Repay, Withdraw \\ 
                  amountETH & Borrow, Deposit, Repay, Withdraw \\ 
                  borrowRate & Borrow \\ 
                  borrowRateMode & Borrow \\ 
                  liquidator & Liquidation \\ 
                  principalAmount & Liquidation \\ 
                  principalReserve & Liquidation \\ 
                  principalReserveType & Liquidation \\
                  principalAmountUSD & Liquidation \\ 
                  principalAmountETH & Liquidation \\ 
                  collateralAmount & Liquidation \\ 
                  collateralReserve & Liquidation \\ 
                  collateralReserveType & Liquidation \\
                  collateralAmountUSD & Liquidation \\ 
                  collateralAmountETH & Liquidation \\ 
                  priceInUsd & Borrow, Deposit, Repay, Withdraw \\ 
                  version & All \\ 
                  deployment & All \\ 
                   \hline
                \end{tabularx}
                \label{tab:rawTransactionFeatures}
            \end{table}

            The user history features are features created by user. For each user, cumulative calculations of the following quantities are created: seconds since first transaction, seconds since last transaction, the count of each transaction types a user has made, the sum and average amount of each transaction type a user has made (in the native amount, Dollars, and Ethereum). The full list of these features can be found in \cref{tab:userFeatures}.

            \begin{table}[ht]
                \centering
                \caption{All user-level features engineered to represent a user's transaction history prior to each transaction. Features with [TYPE] are created identically for each of the five main transaction types (borrow, deposit, liquidation, repay, and withdraw).}
                \rowcolors{1}{}{lightgray!25}
                \begin{tabularx}{\linewidth}{XX}
                  \hline
                Feature Name & Description\\ 
                  \hline
                  userReserveMode & The most common coin used by this user.\\ 
                  userCoinTypeMode & The most common coin type (stable or non-stable) used by this user.\\ 
                  userIsNew & Whether this is the user's first transaction.\\ 
                  userSecondsSinceFirstTransaction & How many seconds have passed since this user's first transaction. \\ 
                  userSecondsSincePreviousTransaction & How many seconds have passed since this user's previous transaction. \\ 
                  userCollateralCount & How many collateral transactions this user has made in the past. \\ 
                  userSwapCount & How many Swap transactions this user has made in the past. \\
                  user[TYPE]Count & How many [TYPE] transactions this user has made in the past. \\ 
                  user[TYPE]Sum & The total amount of the coin involved in this transaction that this user has used in their past [TYPE] transactions. \\ 
                  user[TYPE]AvgAmount & The average amount of the coin involved in this transaction that this user has used per [TYPE] transaction. \\ 
                  user[TYPE]SumUSD & The total value, scaled to USD, of all [TYPE] transactions made by this user in the past. \\ 
                  user[TYPE]AvgAmountUSD & The average value, scaled to USD, per [TYPE] made by this user in the past. \\ 
                  user[TYPE]SumETH & The total value, scaled to Ethereum, of all [TYPE] transactions made by this user in the past.\\ 
                  user[TYPE]AvgAmountETH & The average value, scaled to Ethereum, per [TYPE] transaction made by this user in the past. \\
                  userActiveDaysWeekly & The number of days in the past seven days during which this user has made at least one transaction. \\ 
                  userActiveDaysMonthly & The number of days in the past 30 days during which this user has made at least one transaction. \\ 
                  userActiveDaysYearly & The number of days in the past 365 days during which this user has made at least one transaction. \\ 
                   \hline
                \end{tabularx}
                \label{tab:userFeatures}
            \end{table}
            
            Similarly to user history features, market history features compute metrics for the entire Aave V2 Mainnet market. This is a useful way of approximating a market's relative supply and demand at a given point. For the entire blockchain, features track the number of each type of transaction, the average amount for each type of transaction, and the average and sum amounts of all currencies together in dollars and Ethereum. The full list of these features can be found in \cref{tab:marketFeatures}.
            
            \begin{table}[ht]
                \centering
                \caption{All market-level features engineered to represent an overall market's transaction history before each transaction. Features with [TYPE] are created identically for each of the five main transaction types (borrow, deposit, liquidation, repay, and withdraw).}
                \begin{tabularx}{\linewidth}{XX}
                    \hline
                    Feature Name & Description\\ 
                    \hline 
                    marketCollateralCount & How many collateral transactions that have been made across the whole market in the past. \\ 
                    marketSwapCount & How many Swap transactions that have been made across the whole market in the past. \\
                    market[TYPE]Count & How many [TYPE] transactions have been made across the whole market in the past. \\ 
                    market[TYPE]AvgAmount & The average amount of the coin involved in this transaction that users in this market have used in [TYPE] transactions in the past.\\ 
                    market[TYPE]Sum & The total amount of the coin involved in this transaction that this users in this market have used across all past [TYPE] transactions.\\ 
                    market[TYPE]AvgAmountUSD & The average value, scaled to USD, per [TYPE] made by users in this market in the past.\\ 
                    market[TYPE]SumUSD & The total value, scaled to USD, of all [TYPE] transactions made by users in this market in the past.\\ 
                    market[TYPE]AvgAmountETH & The average value, scaled to ETH, per [TYPE] made by users in this market in the past.\\ 
                    market[TYPE]SumETH & The total value, scaled to ETH, of all [TYPE] transactions made by users in this market in the past.\\ 
                    \hline
                \end{tabularx}
                \label{tab:marketFeatures}
            \end{table}

            Time features were computed for every observation in the raw data. The `timestamp` variable represents the POSIX time (time since January 1, 1970, 00:00:00 UTC) in seconds. Circular representations of these time features were also created to capture the cyclical nature of time, such as the closeness of 11:59 PM and 12:00 AM. This resulted in three additional features per time interval: the value, sine, and cosine. The full list of time features can be found in \cref{tab:timeFeatures}.

            \begin{table}[ht]
                \centering
                \caption{Time-based features created for each transaction in the dataset. Features prefixed with ``sin[cos]'' represent two separate features, one starting with ``sin'' and one starting with ``cos''.}
                \rowcolors{1}{}{lightgray!25}
                \begin{tabularx}{\linewidth}{XX}
                  \hline
                Feature Name & Description \\ 
                  \hline
                  timeOfDay & A numeric value between 0 and 24 representing the time of day at which the transaction took place.\\ 
                  dayOfWeek & The day of the week (1-7) during which the transaction took place.\\ 
                  dayOfMonth & The day of the month (1-31) during which the transaction took place.\\ 
                  dayOfYear & The day of the year (1-365) during which the transaction took place.\\ 
                  quarter & The quarter (1-4) during which the transaction took place.\\ 
                  dayOfQuarter & The day of the quarter (1-95) during which the transaction took place.\\ 
                  sin[cos]TimeOfDay & The time of day transformed by the sine [cosine] function.\\ 
                  sin[cos]DayOfWeek & The day of the week transformed by the sine [cosine] function. \\ 
                  sin[cos]DayOfMonth & The day of the month transformed by the sine [cosine] function.\\ 
                  sin[cos]DayOfQuarter & The day of the quarter transformed by the sine [cosine] function. \\ 
                  sin[cos]DayOfYear & The day of the year transformed by the sine [cosine] function. \\ 
                  sin[cos]Quarter & The quarter transformed by the sine [cosine] function. \\ 
                  isWeekend & A boolean flag (0 or 1) representing whether this transaction occurred on a weekend (0 if no, 1 if yes).\\ 
                   \hline
                \end{tabularx}
                \label{tab:timeFeatures}
            \end{table}

            The data set created contains 128 features in total, of which 106 are constructed from the base features.

\end{document}